# Temporally symmetry-broken metasurfaces for ultrafast resonance creation and annihilation

*Andreas Aigner[1,\*], Thomas Possmayer[1,\*], Thomas Weber[1],*

*Leonardo de S. Menezes[1,2], Stefan A. Maier[3,4], and Andreas Tittl[1,+]*

1) Chair in Hybrid Nanosystems, Faculty of Physics, Ludwig-Maximilians-University Munich, 80539 Munich, Germany.

2) Departamento de Física, Universidade Federal de Pernambuco, 50670-901 Recife-PE, Brazil.

3) School of Physics and Astronomy, Monash University, Clayton, Victoria 3800, Australia.

4) Department of Physics, Imperial College London, London SW7 2AZ, United Kingdom.

[\*] authors contributed equally

[+] andreas.tittl@physik.uni-muenchen.de

## Abstract

Active metasurfaces, compact platforms for nanoscale light manipulation, are transforming technologies like holography, quantum cryptography, and optical computing. Despite their versatility, tunability in metasurfaces has mainly relied on shifting the resonance wavelength or increasing material losses to spectrally detune or quench resonant modes, respectively. However, both methods face fundamental limitations, such as limited Q-factor and near-field enhancement control and the inability to achieve resonance on/off switching by completely coupling and decoupling the mode from the far-field. Here, we demonstrate temporal symmetry-breaking in metasurfaces via ultrafast optical pumping, marking the first experimental realization of radiative loss-driven resonance creation, annihilation, broadening, and sharpening. We introduce restored symmetry-protected bound states in the continuum as a new concept which are central to the realization of temporal symmetry-breaking. These states arise in metasurfaces with geometrically asymmetric unit cells, where the total dipole moment, composed of two antisymmetric dipoles, cancels out. Mie-resonant optical absorption within specific regions of the unit cell locally modifies the refractive index, disrupting the balance between the two dipole moments. A total dipole moment is thereby created or annihilated, and consequently, the radiative loss is tuned. This enables full control over coupling to incoming light, allowing precise adjustment of the resonance linewidth, near-field enhancement, and resonance amplitude. Our work establishes radiative loss-based active metasurfaces with potential applications ranging from high-speed optical and quantum communications to time-crystals and photonic circuits.

**Introduction**

Active nanophotonics is a rapidly advancing field that offers promising solutions for many emerging technologies like holography, quantum cryptography, and optical computing[1,2,3,4]. In particular, active metasurfaces, two-dimensional arrays of subwavelength-spaced nanoresonators, have emerged as a powerful tool for manipulating and confining light[5,6,7]. They have been successfully applied in beam steering[8,9,10,11], optical switching[12,13], holography[14], adjustable lenses[14,15], tunable sensors[16,17], programmable surfaces[18,19,20], and active chiral[21] and polarization[22] filters. In general, tunability of a resonant system, as shown in **Figure 1a**, is achieved by altering one or more of the fundamental resonance parameters: resonance wavelength $\omega_0$, intrinsic loss $\gamma_{\text{int}}$, and radiative loss $\gamma_{\text{rad}}$, with the literature predominantly focusing on $\omega_0$[23,24,25,26,27] and $\gamma_{\text{int}}$[28,29,30,31,32]. Tuning $\omega_0$ shifts the resonance spectrally, while tuning $\gamma_{\text{int}}$ dampens the resonant mode, both resulting in changes to the amplitude at specific wavelengths. The idealized cases of these two tuning methods are illustrated in **Figures 1b** and **1c**, respectively. In reality, however, these parameters are intertwined due to Kramers-Kronig[33] relation linking the real ($n$) and imaginary part ($k$) of the refractive index: $\omega_0$ is primarily influenced by $n$, while $\gamma_{\text{int}}$ by $k$. Although both tuning methods have achieved remarkable results they face inherent limitations. These limitations become evident when examining the far-field response of a single photonic mode, here described by the Lorentzian transmission coefficient[34] (**Figure 1a**):

$$t(\omega) = 1 - \frac{\gamma_{\text{rad}}}{i(\omega - \omega_0) + \gamma_{\text{rad}} + \gamma_{\text{int}}} \qquad (1)$$

It is well known from literature that both $\omega_0$ and $\gamma_{\text{int}}$ can influence the modes' amplitude, linewidth and, with it, the local field enhancement. However, neither can truly turn the mode "on" or "off" by changing the fraction term in Equation 1 from zero to a finite value or vice versa. Such control requires coupling (or decoupling) of the mode to the radiation continuum, effectively toggling the mode between "bright" and "dark" states, which is mediated by the third resonance parameter, $\gamma_{\text{rad}}$. A $\gamma_{\text{rad}}$ of zero means a fully decoupled mode from the far-field, rendering it "off" while $\gamma_{\text{rad}} > 0$ switches the mode "on", both sketched in **Figure 1d**. Furthermore, $\gamma_{\text{rad}}$ directly governs the resonance amplitude, Q-factor, and local field enhancement, making it critical for precise control of metasurface functionalities. Despite its potential, achieving active control of $\gamma_{\text{rad}}$ remains challenging. Simple refractive index modifications yield only minor changes in $\gamma_{\text{rad}}$, far from the complete suppression or enhancement required for full tunability close to $\gamma_{\text{rad}} = 0$. Even in passive metasurfaces, controlling $\gamma_{\text{rad}}$ has only recently become feasible with the introduction of symmetry-protected bound states in the continuum (SP-BICs)[35,36]. These states enable passive $\gamma_{\text{rad}}$ control by breaking the geometric symmetry in a metasurface unit cell.[37] However, first active SP-BICs have focused on tuning $\omega_0$[38,39,40,21,41,42] or $\gamma_{\text{int}}$[16,43,38,32,44], without fully exploiting the unique ability of SP-BICs to adjust $\gamma_{\text{rad}}$. Initial efforts to tune $\gamma_{\text{rad}}$ have been hindered by significant intrinsic losses[16,45] and could only change the Q-factor to a minor degree. Thus, the ability to arbitrarily tune radiative loss to increase or decrease the Q-factor and achieve on/off switching of resonances has not yet been achieved, particularly for low-loss systems.

To address these challenges, we present a novel experimental approach demonstrating temporal symmetry-breaking and $\gamma_{\text{rad}}$ tuning in metasurfaces by ultrafast optical pumping (**Figure 1e**). This enables the creation and annihilation of high Q-factor resonances on a femtosecond timescale, modulating the resonator material through photo-excited charge carriers as a first proof of concept. The temporal tuning of $\gamma_{rad}$ is made possible by leveraging the concept of *restored* SP-BIC (RSP-BICs), which are, for the first time, experimentally introduced in this work. Here, the structural symmetry within a unit cell is broken, but for light at a specific wavelength, the system behaves symmetrically with canceling anti-parallel dipoles, and hence $\gamma_{rad}$ converges to zero. In our work, the unit cells consist of two rods of different lengths and widths (**Figure 1f**) that exhibit equal dipole moments at the RSP-BIC wavelength. However, for other wavelengths and polarizations each has a set of unique Mie modes. This enables us to lower the refractive index $n$ in only one of the two rods which we photoexcite selectively by resonantly pumping its Mie mode. The change in $n$ alters the ratio of the individual dipole moments (**Figure 1g**), modifying their asymmetry and $\gamma_{rad}$ of the metasurface. Notably, small changes of $n$ are sufficient to break the symmetry[46]. In transient absorption experiments, we achieve ultrafast control of $\gamma_{rad}$ near the RSP-BIC condition in four ways: We can sharpen, broaden, create, and annihilate resonances depending on the system's geometry and pump fluence, while the effect on $\gamma_{int}$ is negligible.

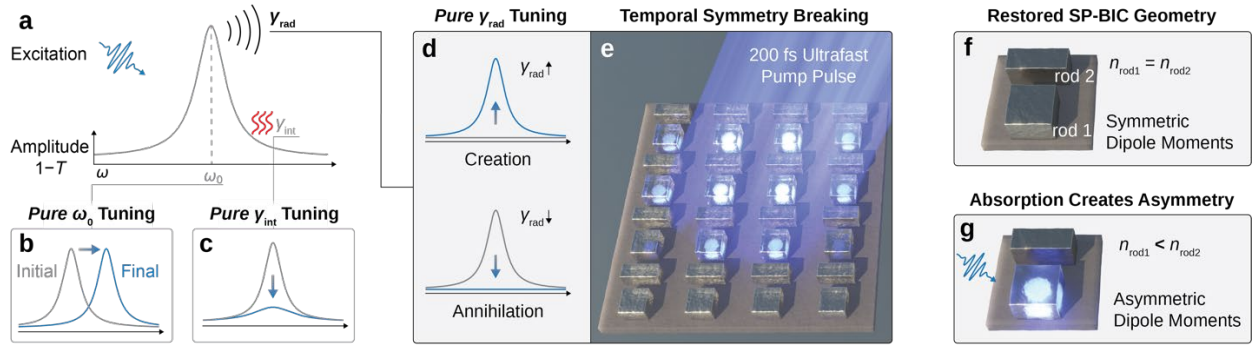

**Figure 1: Active tuning mechanisms and ultrafast radiative loss control via selective pumping. (a)** Illustration of an initial resonant mode before excitation using $1 - t$ As indicated, it is defined by three parameters: $\omega_0$, $\gamma_{int}$, and $\gamma_{rad}$. **(b-d)** Active tuning approaches for $\omega_0$, $\gamma_{int}$, and $\gamma_{rad}$, respectively, with the gray curve as the initial resonance and the blue curve as the tuned one. **(b)** Tuning $\omega_0$ results in a shifted resonance profile while the amplitude and spectral width remain unchanged. **(c)** Tuning $\gamma_{int}$ quenches the amplitude of the resonance without fully eliminating it. **(d)** Tuning $\gamma_{rad}$ allows the mode to be created as $\gamma_{rad}$ increases from zero and fully annihilated as $\gamma_{rad}$ decreases to zero. **(e)** Illustration of the temporal symmetry-breaking metasurface, selectively pumped with a 200 fs pulse resonant with a Mie mode in only one of two rods per unit cell indicated by the glowing areas. **(f)** The metasurface initially exhibits an RSP-BIC, where the dipole moments of both rods are symmetric due to matching refractive indices ($n_{rod1} = n_{rod2}$). **(g)** After resonant absorption by the pump pulse, the refractive index in rod 1 decreases ($n_{rod1} < n_{rod2}$), resulting in asymmetric dipole moments and the emergence of the SP-BIC mode ($\gamma_{rad} \neq 0$).

## Restored Symmetry-Protected Bound States in the Continuum

Conventional SP-BICs rely on in-plane geometric inversion symmetry within the unit cell to suppress coupling to radiative channels. Fundamentally, however, coupling is prevented if the effective dipole moment of the unit cell is zero. Remarkably, we demonstrate that it is possible to break the geometric symmetry of the resonators while maintaining a zero effective dipole moment: the restored SP-BIC (RSP-BIC) condition with $\gamma_{\text{rad}} = 0$.

We use crystalline silicon as resonator material due to its optical tunability[47,48,45] and fairly low losses at the target wavelength of 800 nm. The nanoresonators with a height of 115 nm were fabricated on a sapphire substrate and encapsulated with silicon dioxide ($SiO_2$). Our chosen geometry comprises two aligned dipolar rods within a $420 \times 420$ nm$^2$ unit cell (**Figure 2a**). When their respective lengths and widths match, the system exhibits symmetry protection: the opposing dipole moments cancel ($p_{\text{tot}} = 0$), and the mode is decoupled from the far-field, resulting in a vanishing $\gamma_{\text{rad}}$ (**Figure 2a**). Breaking the symmetry by increasing the width $w_1$ of the first resonator increases its dipole moment. This yields a net asymmetry of the combined structure and a quasi-BIC forms with $p_{\text{tot}} > 0$ (**Figure 2b**). Next, we increase the length $l_2$ of the second resonator increasing its dipole moment as well. At a specific combination of $w_1$ and $l_2$, the dipole moments match again ($p_{\text{tot}} = 0$), restoring the optical symmetry despite the broken geometric symmetry. This is the RSP-BIC condition, where $\gamma_{\text{rad}}$ returns to zero, and the radiative Q-factor diverges (**Figure 2c**).

In simulations (see Methods), the initial resonator lengths and widths are set to 175 nm and 95 nm, respectively. **Figure 2d** shows the emergence of the SP-BIC mode around 770 nm when $w_1$ increases. The fitted $\gamma_{\text{rad}}$ in **Figure 2e** (TCMT model in **Supplementary Note 1**) converges to zero around the symmetric case, which is typical for SP-BICs. **Figure S1** demonstrates the typical relation for SP-BICs of $Q_{\text{rad}} \propto 1/\alpha^2$ with the asymmetry factor $\alpha$. Continuing with the asymmetric case ($w_1 = 185$ nm), increasing $l_2$ (right panel of **Figure 2d**) causes the mode to sharpen and eventually disappear at $l_2 = 216$ nm, which corresponds to the RSP-BIC condition with $\gamma_{\text{rad}} = 0$ (right panel of **Figure 2e**). For even larger $l_2$, the mode reappears. Again, **Figure S1** finds good agreement with $Q_{\text{rad}} \propto 1/\alpha^2$ before and after the RSP-BIC condition while **Figure S2** reveals an identical mode profile for all $l_2$, confirming the SP-BIC nature of both branches.

Based on these numerical results, we fabricated nanoresonators that match the simulated designs (see Methods and workflow in **Figure S3**). Scanning electron microscopy (SEM) images of the unit cells before $SiO_2$ encapsulation are shown in **Figure 2f** at the true SP-BIC, quasi SP-BIC, and RSP-BIC conditions. To ensure a continuous transition between these states, we design two gradient metasurfaces[49] mimicking the numerical results shown in **Figure 1d**, each gradient with a size of $100 \times 50$ µm$^2$. The first continuously increases $w_1$ from 95 nm to 185 nm, while the second uses $w_1 = 185$ nm and increases $l_2$ from 175 nm to 275 nm. A true color optical image is shown in **Figure 2g**, where the gradual color change indicates the smooth variation of $w_1$ and $l_2$ (SEM images in **Figure S4**). Experimental transmittance spectra in **Figure 2h**, extracted along the x-axis (see Methods and **Figure S5** for line spectra), confirm the presence of the same SP-BIC mode as seen in **Figure 2d**. For the gradient on the right, the RSP-BIC condition is evident in **Figure 2i** when $\gamma_{\text{rad}}$ converges to zero before reappearing. The seamless tracing of this mode back to the conventional SP-BIC, both in simulations and experiments, combined with the $1/\alpha^2$ dependency of $Q_{\text{rad}}$, strongly supports the symmetry-protected nature of the RSP-BIC condition.

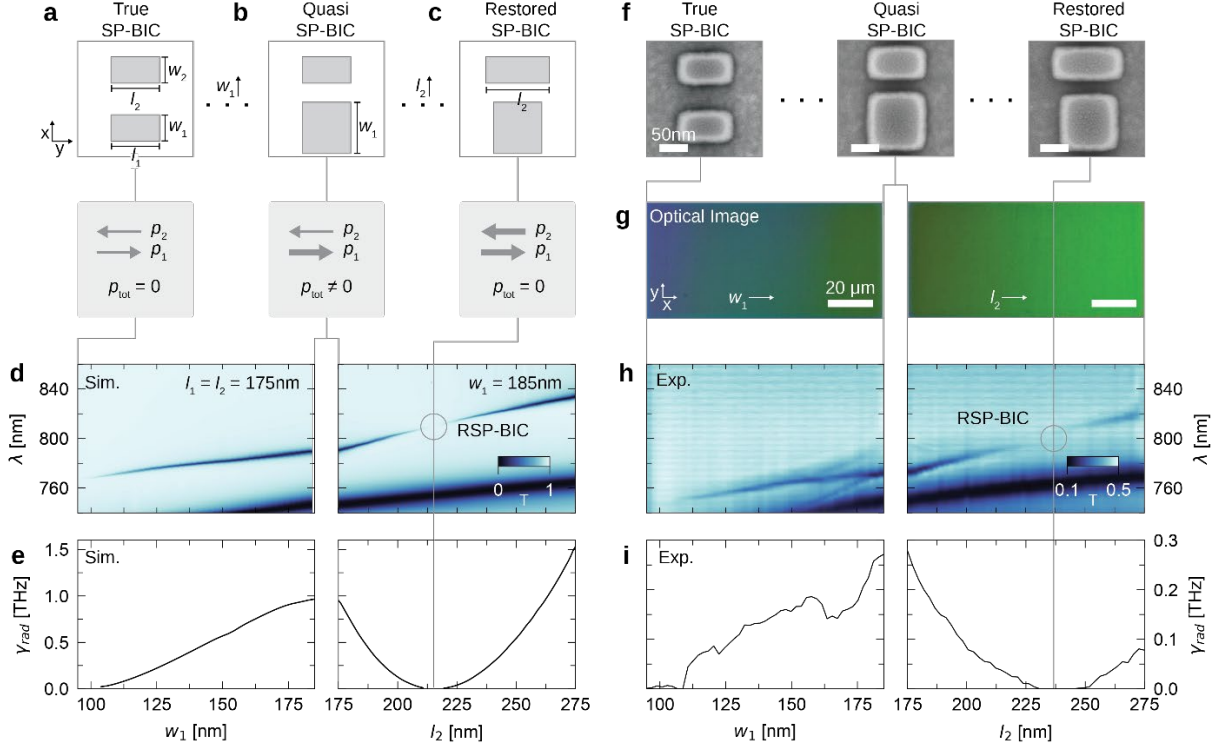

**Figure 2: Restored SP-BIC principle and experimental verification. (a)** Sketch of the unit cell geometry, consisting of two crystalline silicon rods with lengths $l_1$, $l_2$ and widths $w_1$, $w_2$, respectively. The dipole moments $p_1$ and $p_2$ along the x-axis are equal, resulting in a total dipole moment $p_{tot} = 0$ for an out-of-phase mode, indicating a true SP-BIC condition. **(b)** When $w_1$ is increased, the symmetry is broken (quasi-BIC), and $p_{tot} \neq 0$, allowing the mode to couple to the far-field. **(c)** Increasing $l_2$ restores the symmetry, returning the system to $p_{tot} = 0$, the RSP-BIC condition. **(d)** Numerical transmittance spectra of the SP-BIC mode as $w_1$ is varied from 95 to 185 nm (left). On the right, tuning $l_2$ from 175 to 275 nm for fixed $w_1 = 185$ nm sharpens the mode until it disappears at the RSP-BIC (marked by the gray circle). **(e)** $\gamma_{rad}$ obtained from TCMT fitting, converging to zero at the SP-BIC and RSP-BIC conditions. **(f)** SEM images of the crystalline silicon metasurface corresponding to the cases shown in (a-c). **(g)** Optical image of the two gradient metasurfaces, with $w_1$ gradient on the left and $l_1$ gradient on the right. **(h)** Experimental spectra matching the numerical results in (d). **(i)** Fitted experimental data for $\gamma_{rad}$, corresponding to the results in (e).

## Selective Optical Pumping by Dissimilar Mie Resonances

After experimentally verifying the RSP-BIC condition with $\gamma_{rad} = 0$ in a highly asymmetric unit cell, we now investigate additional optical modes of the system that could be used for selective optical pumping. The transmittance spectra of the metasurfaces at the RSP-BIC condition are shown in **Figure 3a** for both x- and y-polarized light, aligned along and normal to the rods, respectively. While the SP-BIC mode is not visible due to $\gamma_{rad} = 0$, three distinct dips are visible, which we attribute to Mie modes: two for y-polarized light at 720 and 745 nm, labeled Mie 1 and Mie 2, respectively, and one for x-polarized light at 764 nm labeled Mie 3. Based on multipole decompositions (**Supplementary Note 2**) shown in **Figure S6**, we can assign an electric dipole-like behavior to Mie 1, while Mie 2 and Mie 3 are magnetic dipole-like modes.

As the active tunability is based on lowering $n$ via photoexcitation, we compare the loss density distribution within both rods. **Figure 3b** shows the simulated loss density profile for all three Mie modes cut at a height of 30 nm. Strikingly, we observe higher loss densities in Rod 1 for all three modes, with a quantitative comparison of loss per volume in **Figure 3c**. Mie 1 exhibits a particularly strong imbalance with a 3.5-fold higher absorption per volume in Rod 1 than in Rod 2. As a high absorption imbalance yields a high $n$ imbalance, we select Mie 1 as the ideal mode for efficient $\gamma_{\text{rad}}$ tuning.

Our tuning approach is based on above-bandgap pumping of silicon, sketched in **Figure 3d.** This modifies the material's polarizability, lowering the refractive index[50] from $n_{\text{Si},0}$ to an absorption dependent $n_{\text{Si,Pump}}$. This decrease occurs on the timescale of the pump pulse (here 200 fs), while its recovery is mainly defined by carrier recombination through three mechanisms: surface, trap, and Auger recombination. Due to our structure's high surface-to-volume ratio and defect density caused by nanofabrication, the recombination is expected to happen significantly faster than in bulk silicon.

We use a geometry with $l_2 \approx 236$ nm to ensure the SP-BIC mode is visible at 799 nm, while still the same Mie modes as in the RSP-BIC case are present, see **Figure S7**. To excite Mie 1, we pump the sample with 720 nm and a fluence of 100 µJ/cm$^2$ (pump-probe setup sketched in **Figure S8**). A broadband probe pulse monitors changes in the transmission at variable delay times. Investigating non-Mie-resonant pumping first, we pump the sample with x-polarized light at 720 nm (**Figure 3a**). At a delay time $t = 0$ s, the SP-BIC mode shifts by 3.9 nm (7.6 meV, **Figure 3e**), due to the overall change of $n$. The spectral shift decays exponentially with a 15 ps decay constant without a significant change of the SP-BIC amplitude. This indicates that both rods were pumped equally, without affecting $\gamma_{\text{rad}}$.

Next, we switch to Mie-resonant enhanced absorption using a 720 nm y-polarized pump (see **Figure S9** for a pump wavelength sweep). In the resulting time trace (**Figure 3f**), a much stronger spectral shift of 9.2 nm (18 meV, equivalent to 2.2 FWHMs) is observed. This stronger shift confirms resonant absorption, while the increase in resonance amplitude suggests a change in $\gamma_{\text{rad}}$.

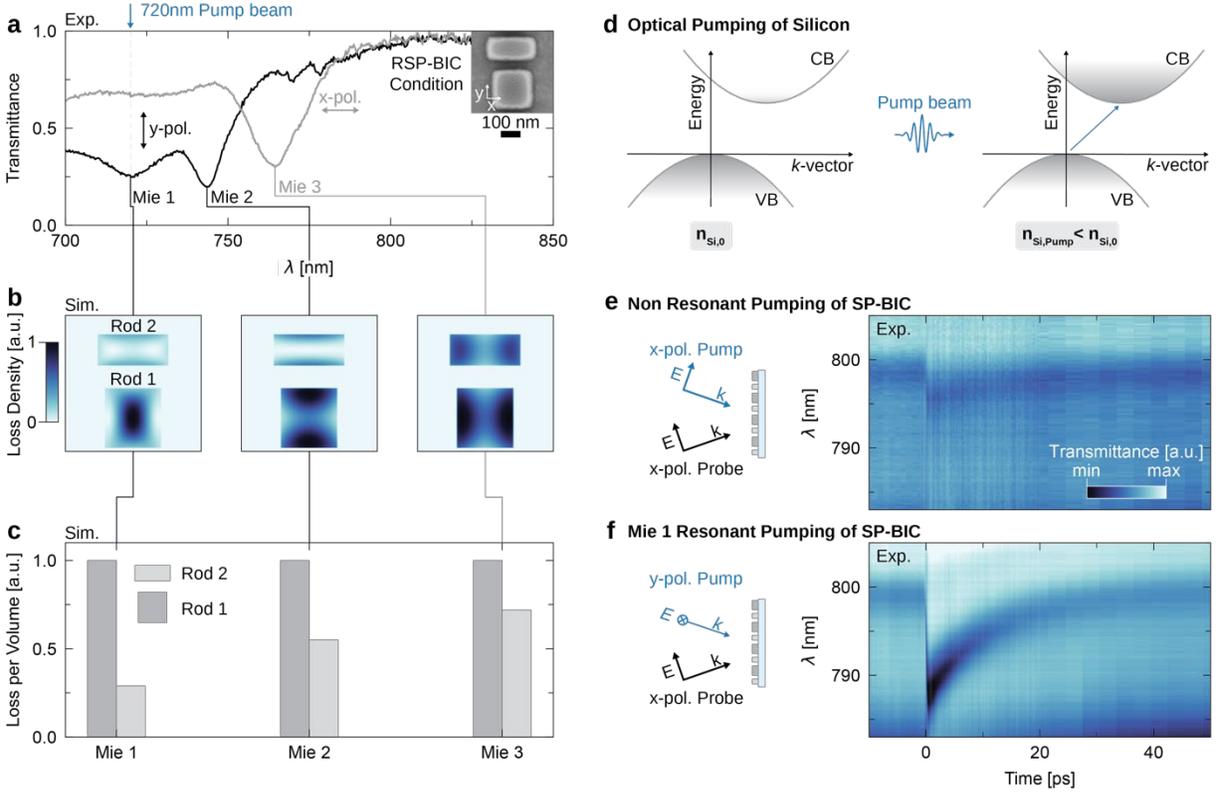

**Figure 3: Mie modes and selective resonant pumping.** (a) Experimental transmittance spectrum at the RSP-BIC condition with $l_2 \approx 226$ nm for x- and y-polarized light, revealing two and one Mie modes, respectively (labeled Mie 1, Mie 2, and Mie 3). (b) The unit cells' normalized loss density map for the three Mie modes shows distinct mode profiles and dissimilar losses in both rods at a cutting plane of $z = 30$ nm. (c) Comparison of normalized loss per volume between the two rods for each Mie mode. Mie 1 exhibits the highest ratio between Rod 1 and Rod 2, indicating the strongest selective absorption. (d) Sketch of the optical pumping principle: Above-bandgap excitation of carriers from the conduction band (CB) to the valence band (VB) generates free electrons and holes, altering the polarizability and the refractive index. (e) Pump-probe spectral time trace (720 nm pump with a fluence of 100 µJ/cm²) for a metasurface with $l_2 = 236$ nm, where both pump and probe pulses are x-polarized. The SP-BIC at 799 nm shifts 3.9 nm and exponentially returns to the initial wavelength with a time constant of 15 ps, with minimal changes in resonance amplitude. (f) Same configuration as in (e) but with the pump pulses y-polarized, resonantly pumping Mie 1. The SP-BIC shifts 9.2 nm, and an apparent change in the resonance amplitude is visible.

## Ultrafast Tuning of Radiative Losses

The pump beam resonantly excites Rod 1 and, therefore, reduces its dipole moment $p_1$ relative to $p_2$ of Rod 2. Depending on the geometry of the unit cell, this can either increase or decrease $\gamma_{\text{rad}}$: in areas of the gradient where $p_1 > p_2$, $\gamma_{\text{rad}}$ decreases, as the difference between the dipole moments reduces. Conversely, in areas where initially $p_1 < p_2$, $\gamma_{\text{rad}}$ increases. To illustrate this behavior, the transmittance around the RSP-BIC for an $l_2$ sweep is sketched in gray in **Figure 4a**. It features two distinct regions divided by the line where $p_1 = p_2$. Upon pumping, $p_1$ is lowered and this line shifts, as indicated by the transmittance signal in blue. To demonstrate the possible switching effects, we select four distinct points along the $l_2$ sweep, marked by gray cuts. Each cut represents a different initial relationship between $p_1$ and $p_2$ and is further discussed on the right of **Figure 4**: $p_1 < p_2$ in **Figure 4b**, $p_1 = p_2$ in **Figure 4c** (RSP-BIC condition), $p_1$ slightly greater than $p_2$ in **Figure 4d**, and $p_1 > p_2$ in **Figure 4e**.

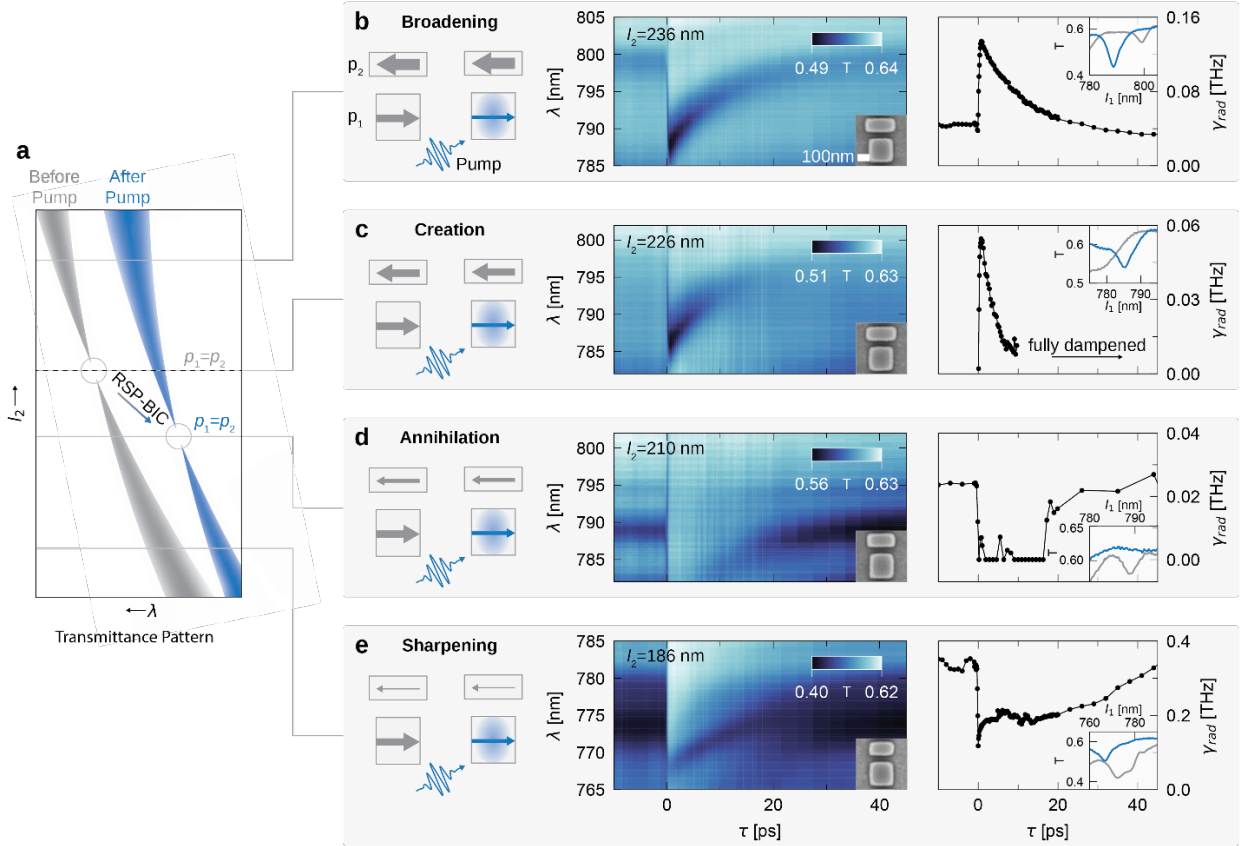

**Figure 4: Temporal radiative loss tuning around the RSP-BIC condition. (a)** Illustration of the SP-BIC mode around the RSP-BIC condition for an $l_2$ sweep before (gray) and after (blue) pumping of the structure. Note that the RSP-BIC condition marked with circles shifts to smaller $l_2$ after pumping. A dashed line marks where $p_1 = p_2$. Four key positions are highlighted with gray cuts where resonances are broadened (b), created (c), annihilated (d), or sharpened (e), all dependent on the initial ratio of $p_1$ to $p_2$. **(b)** Transmittance time evolution with a 100 µJ/cm² pump pulse at 720 nm of the gradient position with $l_2 = 236$ nm, where $p_1 < p_2$, showing an increase in resonance amplitude and the radiative loss $\gamma_{rad}$. On the top right transmittance spectra at $t = 0$ ps and $t = 1$ ps. **(c)** Time evolution with $l_2 = 226$ nm, where $p_1 = p_2$ for the initial structure. After pumping, a mode is created, which quickly decays after 10 ps. This is also reflected in the $\gamma_{rad}$ fit, which increases from 0 to 0.058 THz before exponentially decaying. The fit is restricted to the time interval where the resonant mode is visible. In the top right corner, the transmittance spectrum at $t = 0$ ps and $t = 1$ ps. **(d)** Time evolution with $l_2 = 210$ nm, where $p_1 > p_2$, showing the annihilation of the BIC mode, which reappears after approximately 20 ps. **(e)** Time evolution with $l_2 = 186$ nm, where $p_1 > p_2$, resulting in a decrease of $\gamma_{rad}$ and sharpening of the mode.

The first case shown in **Figure 4b** with $l_2 = 236$ nm and $p_1 < p_2$ features a visible SP-BIC mode around 799 nm. Upon pumping, the difference between the two dipole moments further increases. Hence, the mode *broadens* and the resonance amplitude as well as $\gamma_{rad}$ (from 0.04 to 0.14 THz) significantly increases before gradually returning within 20 ps. The total Q-factor changes by 25% from 400 to around 300, which is plotted in **Figure S10a**. The second case at the RSP-BIC condition (**Figure 4c**) with $l_2 = 226$ nm and $p_1 = p_2$, initially features no resonance. After pumping the balance shifts to $p_1 < p_2$ and the resonance is *created*. The newly formed mode at 787 nm features $\gamma_{\text{rad}}$ of up to 0.058 THz, which decays back to values close to zero within 10 ps, when the RSP-BIC condition is restored. Note that for times below 0 ps and above 10 ps, the TCMT fit is not conclusive due to the absence of the mode. For the third case in **Figure 4d**, with $l_2 = 210$ nm and $p_1 > p_2$, the pump pulse reduces the dipole imbalance, effectively restoring the symmetry with $p_1 = p_2$. Thus, the initial resonance is *annihilated* as $\gamma_{\text{rad}}$ drops from 0.025 to 0 THz. The resonance reappears after 20 ps as the system recovers. Finally, the fourth case shown in **Figure 4e**, with $l_2 = 186$ nm and $p_1$ much larger than $p_2$, the pump reduces $p_1$, but $p_1 > p_2$ still holds. Rather than restoring the symmetry, the dipole imbalance is reduced but remains nonzero. This causes a *sharpening* of the resonance, with a decrease in amplitude, while $\gamma_{\text{rad}}$ drops from 0.35 to 0.12 THz, and the total Q-factor increases by 150% from around 100 to 250 (**Figure S10b**).

Next, we investigate two essential tuning parameters: onset speed and continuous tunability. **Figure 5a** shows a transmittance map for short time delays and $l_2 = 236$ nm. The onset occurs nearly instantaneously, here limited to 300 fs by the pump and probe pulse widths. During the rise, $\gamma_{\text{int}}$ increases significantly (**Figure 5b**), which we attribute to two main factors: nondegenerate two-photon absorption and the spectral shift of the resonant mode. The latter is caused by the cavity's resonance lifetime (around 1 ps), which is longer than the 300 fs rise time. This causes the cavity to change faster than the light within decays. Contrary to $\gamma_{\text{int}}$, $\gamma_{\text{rad}}$ decays much slower, allowing optimal resonance performance starting at 300 fs. **Figure S11** displays longer time delays.

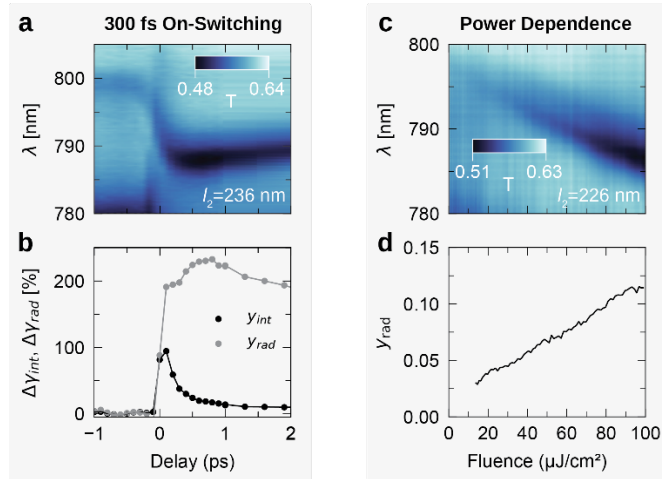

**Figure 5: On-switching dynamics and power dependence. (a)** Close-up of the transmittance time evolution for $l_2 = 236$ nm. **(b)** Corresponding changes in intrinsic loss $\gamma_{\text{int}}$ and $\gamma_{\text{rad}}$, showing a sharp increase for $\gamma_{\text{int}}$ by approximately 100% upon pump arrival, followed by a rapid drop to approximately 14% above pre-pump values within 1 ps while $\gamma_{\text{rad}}$ remains at high values. **(c)** Pump-fluence-dependent transmittance spectrum for $l_2 = 226$ nm (case of resonance creation), showing a gradual increase in resonance amplitude with increasing pump fluence. **(d)** Corresponding $\gamma_{\text{rad}}$ values derived from fits to the data in (c), demonstrating a continuous increase in $\gamma_{\text{rad}}$ with fluence, reaching up to 0.11 THz.

To illustrate continuous tunability, **Figure 5c** presents a transmittance map at 1 ps with varying fluence from 0 to 100 µJ/cm² for $l_2 = 226$ nm corresponding to resonance creation. As the pump fluence increases, a gradual spectral shift and increase in modulation strength are observed. This demonstrates continuous tunability of $\gamma_{\text{rad}}$ from near zero up to 0.11 THz (equivalent to a change of $Q_{\text{rad}}$ from infinity down to 3300), as shown in **Figure 5d**. The fluence-dependent decay time is shown in **Figure S12**.

**Conclusion**

In this work, we present the first experimental demonstration of radiative loss tuning in metasurfaces through temporal symmetry-breaking, allowing to couple or decouple a mode from the far-field. Central to this innovation are RSP-BICs, which enable photonic symmetry in systems with broken geometric symmetry. This allows selective resonant pumping, providing precise control over asymmetry and radiative loss within the SP-BIC framework. We experimentally achieve resonance broadening ($\Delta Q = 100$, a change of 25%), sharpening ($\Delta Q = 150$, a change of 150%), and resonance creation and annihilation on 300 fs timescales. We can continuously tune $\gamma_{\text{rad}}$ from 0 to 0.11 THz (divergent $Q_{\text{rad}}$ down to 3300) by the pump fluence, with an estimated change of $\gamma_{\text{int}}$ by only approximately 14%. This underlines the high selectivity of our method.

The ability of $\gamma_{\text{rad}}$ to precisely control cavity parameters offers many new possibilities for studying light-matter coupling effects, such as quantum emission or polaritonic coupling. This is in contrast to conventional $\omega_0$ and $\gamma_{\text{int}}$ tuning methods, which struggle to dynamically control local field enhancement or spectral width without introducing significant parasitic losses. Furthermore, the demonstrated rapid onset can induce significant time-varying dispersion which has been theoretically predicted.[51] This ultrafast change in the dispersive properties of the system can further be used for time-crystals,[52,53,54] which so far had to rely on the optical modulation of intrinsic resonances, like the epsilon-near-zero modes, to achieve significant effects. Our technique is not limited to silicon-based systems and can be applied to various low-loss dielectrics that can be optically pumped, such as gallium arsenide.[42] Even non-volatile phase-change materials can be used, which we demonstrate numerically by using $Sb_2S_3$ in **Supplementary Note 3** and **Figure S13**. Based on these results, $Sb_2S_3$ can enable programmable metasurfaces with tunable $\gamma_{\text{rad}}$ that can be optically written and erased, providing a platform for reconfigurable photonic devices. Additionally, our method could be extended to faster switching mechanisms based on nonlinear optical effects, such as the Kerr effect,[55,56] enabling even shorter switching times and expanding its utility for ultrafast applications.

**Acknowledgments**

This project was funded by the Deutsche Forschungsgemeinschaft (DFG, German Research Foundation) under grant numbers EXC 2089/1–390776260 (Germany's Excellence Strategy) and TI 1063/1 (Emmy Noether Program), the Bavarian program Solar Energies Go Hybrid (SolTech), and Enabling Quantum Communication and Imaging Applications (EQAP), and the Center for NanoScience (CeNS). Funded by the European Union (ERC, METANEXT, 101078018). Views and opinions expressed are however those of the author(s) only and do not necessarily reflect those of the European Union or the European Research Council Executive Agency. Neither the European Union nor the granting authority can be held responsible for them. S.A.M. additionally acknowledges the Lee-Lucas Chair in Physics.

**Author contributions**

A.A. and A.T. conceived the idea and planned the research. A.A. performed the sample fabrication. A.A. performed the linear measurements, and T.P. performed the time-resolved measurements. A.A. and T.W. conducted the numerical simulations. A.A., T.P., and T.W. contributed to the data processing. A.A., T.P., T.W., and A.T. contributed to the data analysis. S.A.M., L.de S. M., and A.T. supervised the project. All authors contributed to the writing of the paper.

**Competing interests**

The authors declare that they have no competing interests.

**Data and materials availability**

All data are available in the main text or the supplementary materials.

# Supplementary Information:
# Temporally symmetry-broken metasurfaces for ultrafast resonance creation and annihilation

*Andreas Aigner[1,*], Thomas Possmayer[1,*], Thomas Weber[1],*

*Leonardo de S. Menezes[1,2], Stefan A. Maier[3,4], and Andreas Tittl[1,+]*

1) Chair in Hybrid Nanosystems, Faculty of Physics, Ludwig-Maximilians-University Munich, 80539 Munich, Germany.
2) Departamento de Física, Universidade Federal de Pernambuco, 50670-901 Recife-PE, Brazil.
3) School of Physics and Astronomy, Monash University, Clayton, Victoria 3800, Australia.
4) Department of Physics, Imperial College London, London SW7 2AZ, United Kingdom.

[*] authors contributed equally

[+] andreas.tittl@physik.uni-muenchen.de

**Numerical Methods**

We conducted the simulations using CST Studio Suite (Simulia), a commercial finite element solver. The setup included adaptive mesh refinement and periodic boundary conditions in the frequency domain. Crystalline silicon was modeled according to the data provided by Schinke *et al.*[1] For sapphire and $SiO_2$, we applied constant refractive indices of 1.75 and 1.44, respectively, assuming no material losses. The $SiO_2$ layer's height was not considered. Since this layer's thickness is significantly larger than the thickness of the silicon structure, the $SiO_2$/air interface was excluded from the model.

**Sample Fabrication**

As the basis for the sample, we used a commercially available 150 nm crystalline silicon film on a sapphire substrate (Si (100) Epi on R-Plane Sapphire, The Roditi International Corporation Ltd). First, we etched the 150 nm silicon film down to 115 nm using inductively coupled plasma reactive ion etching (ICP-RIE) in a $Cl_2$/Ar gas mixture (PlasmaPro 100 system, Oxford Instruments). Next, we deposited a 40 nm layer of amorphous chromium (Cr) via sputtering (AMOD, Angstrom Engineering Inc.). In the following, we spin-coated 125 nm of electron beam lithography (EBL) resist (CSAR 62, Allresist GmbH) and wrote an inverse pattern of the two rods (eLINE Plus, Raith GmbH) at 30 kV with a 10 μm aperture. The patterned film was developed in an amyl acetate bath, followed by an MIBK (1:9) bath. The Cr layer was then patterned by ICP-RIE using the resist as a mask and a $Cl_2/O_2$ gas mixture. The remaining resist was removed using Microposit Remover 1165 (Microresist GmbH). The patterned Cr was subsequently used as a hard mask for etching the silicon layer with ICP-RIE in a $Cl_2$/Ar gas mixture (PlasmaPro 100 system, Oxford Instruments). The Cr mask was removed with a chromium wet etchant (Chromium Etchant Standard, Merck KGaA). Finally, the sample was encapsulated by spin-coating undoped spin-on-glass (NDG-7000,

Desert Silicon LLC) at 1000 rpm, followed by baking at 150°C for 30 minutes. The fabrication workflow is sketched in **Figure S3**.

**Optical Characterization**

The steady-state transmittance measurements were performed using a confocal microscope (WITec Wissenschaftliche Instrumente und Technologie GmbH) with a white light source (Thorlabs OSL2 Fiber Illuminator), collimated via a collimator, and a high-magnification objective (100x, NA = 0.9 Zeiss AG) to collect the transmitted light which was then measured with a WITec microscope. To reduce the collection area to approximately 1 µm in diameter on the sample surface, an aperture was placed after the collecting objective. This small collection area is essential for reducing the spectral response to small regions within the spatially varying gradient metasurfaces. The entire gradient was then measured using a stepwise rastering of the sample in the x-y plane to create a spectral map, from which we extracted the relevant data.

Time-resolved measurements were performed using a mode-locked Yb:KGW laser (Pharos Ultra II) at a 200 kHz repetition rate pumping an optical parametric amplifier (ORPHEUS-HP, Light Conversion) to generate laser pulses with a tunable wavelength of roughly 190 fs duration. The output was tuned to the wavelength of the pump mode, and split in two using a beamsplitter: One part (pump path) went directly to the sample, while the other (probe path) was focused on a Sapphire Crystal to generate a supercontinuum. After passing a longpass filter to filter out the pump wavelength, and a delay stage to control the time delay between pump and probe, it was recombined with the pump path using a beam splitter. Both pump and probe polarizations could be controlled independently using half-wave plates. The pump and probe beams were condensed on the structure using a 10x, 0.25 NA objective. To achieve a large-area uniform illumination, the pump path contained an additional lens aiming for the back focal plane of the objective. The transmitted supercontinuum was collected with another 10x, 0.25 NA objective and analyzed with a spectrometer (Princeton Instruments Acton SP2300) using a silicon CCD (Princeton Instruments Pixis 100f) and a 300 g/mm grating. The transmitted pump was filtered out using an additional longpass filter in front of the spectrometer.

**Supplementary Note 1: TCMT model**

The following section is based on the work of Fan *et al.*[2] on temporal coupled-mode theory for the Fano resonance in optical resonators[2]. We use a single resonator coupled to two ports, allowing transmission and reflection. The resonance is excited through port 1, with the incoming wave represented as $s_+ = (s_{1+}, 0)^T$, and the output waves in port 1 (reflected wave) and port 2 (transmitted wave) are represented by $s_- = (s_{1-}, s_{2-})^T$, with $s_{1+}$, $s_{1-}$, and $s_{2-}$ as the incoming, reflected, and transmitted wave's amplitude, respectively.

The resonant mode's time-dependent amplitude, denoted as $a(t)$, temporally evolves according to

$$\frac{da(t)}{dt} = (i\omega_0 - \gamma_{tot}) a(t) + \kappa^T s_+$$

where $\omega_0$ is the resonance frequency, $\gamma_{tot} = \gamma_{rad} + \gamma_{int}$ is the total decay rate as the sum of radiative loss $\gamma_{rad}$ and intrinsic loss $\gamma_{int}$. The outgoing waves $s_-$ are related to the incoming waves and the resonator amplitude via

$$s_- = C s_+ + a(t) \kappa.$$

Here, $\kappa = \left(\sqrt{\gamma_{rad}}, \sqrt{\gamma_{rad}}\right)^T$ is the radiative damping rate describing the coupling between the ports and the mode while $C$ represents the nonresonant port-to-port coupling with

$$C = e^{i\varphi} \cdot \begin{pmatrix} r_0 & it_0 \\ it_0 & r_0 \end{pmatrix}.$$

$r_0$ and $t_0$ are the background reflection and transmission, respectively, with $r_0^2 + t_0^2 = 1$, while $\varphi$ represents the global phase.

Using the above formulas and assuming a time-harmonic mode amplitude, $\frac{da(t)}{dt} = i\omega a(t)$, the equation for the resonance's time-dependent amplitude can be written as

$$a(t) = \frac{\kappa^T s_+}{(i\omega - i\omega_0 - \gamma_{tot})}.$$

Assuming unitary transmission for off-resonance wavelengths and using $s_{2-} = \sqrt{\gamma_{rad}} a(t)$, we can substitute $a(t)$ and eventually calculate the transmission coefficient $t(\omega) = \frac{s_{2-}}{s_{1+}}$ as

$$t(\omega) = 1 - \frac{\gamma_{rad}}{i(\omega - \omega_0) + \gamma_{tot}} = 1 - \frac{\gamma_{rad}}{i(\omega - \omega_0) + \gamma_{rad} + \gamma_{int}}.$$

**Supplementary Note 2: Multipole decomposition**

To study the multipolar composition of the present photonic modes in the metasurface, we calculate the scattered power of one unit cell of the metasurface

$$P_{scat} = \frac{1}{2} \sqrt{\frac{\epsilon_0 \epsilon}{\mu_0}} \int |E_{scat}|^2 d\Omega$$

where we decompose the scattered electric fields into several multipolar moments[3]

$$E_{scat}(r) = \frac{k_0^2 e^{ikr}}{4\pi\epsilon_0 r}\left([n \times [p \times n]] + \frac{1}{v}[m \times n] + \frac{ik}{2}[n \times [n \times \hat{Q}_e n]] + \frac{ik}{2v}[n \times \hat{Q}_m n] \right.$$
$$\left. + \frac{k^2}{6}[n \times [n \times \hat{O}_e nn]]\right)$$

with **p** being the electric dipole moment (ED), **m** the magnetic dipole moment (MD), $Q_{e/m}$ the electric/magnetic quadrupole moments (EQ/MQ) and $O_e$ the electric octupole moment (EO). The multipole moments are spatial integrals over the current density

$$J = -i\omega\epsilon_0(\epsilon_r - \epsilon)E$$

where **E** is the electric field inside the resonator with $\epsilon_r$ denoting the relative permittivity of the resonator material and $\epsilon$ the relative permittivity of the environment.

**Supplementary Note 3: $Sb_2S_3$ metasurface**

We numerically investigated $Sb_2S_3$ as a potential alternative material for selective optical pumping around the RSP-BIC condition. Unlike optically pumped silicon, the refractive index change in $Sb_2S_3$ is based on crystallization or amorphization, making it non-volatile and therefore an ideal candidate for reprogrammable or re-adjustable passive devices. Typically, pulsed laser illumination can induce the transition from amorphous to crystalline form, where reaching a critical intensity triggers a phase change.

We employed a comparable geometry as previously used for silicon. As shown in **Figure S13a**, at the RSP-BIC condition, we use a two-rod configuration with varied width and diameter. Although a pure $Sb_2S_3$ resonator would enhance switching performance, we limited the $Sb_2S_3$ layer to 50 nm thickness within a total 160 nm high silicon resonator, as prior experimental studies suggest that thicker films are challenging to switch completely. We selected a resonance wavelength around 1060 nm and used refractive index values for amorphous $Sb_2S_3$ from Delaney et al.[3]. $SiO_2$ was chosen as the substrate, and the unit cell was set to 600 x 600 $nm^2$, with $l_1 = w_1 = 235$ nm, $w_2 = 120$ nm, and $l_2$ swept from 270 to 330 nm, as shown in the transmittance map in **Figure S13b**. The SP-BIC mode clearly sharpens to an RSP-BIC before broadening again, demonstrating the expected RSP-BIC behavior.

In the next step, we assume that selective optical pumping crystallizes the $Sb_2S_3$ layer in Rod 1, shifting the refractive index from 2.829 to 3.413. Under these conditions, the transmittance spectrum across the $l_2$ sweep shows a different pattern, with the RSP-BIC disappearing within the plotted $l_2$ range. Comparing the spectra for $l_2 = 302$ nm, where the RSP-BIC occurred for amorphous $Sb_2S_3$, to the case where $Sb_2S_3$ in Rod 1 is crystalline, we observe that the symmetry is disrupted, and the SP-BIC mode re-emerges at 1072 nm.

**Supplementary Figures**

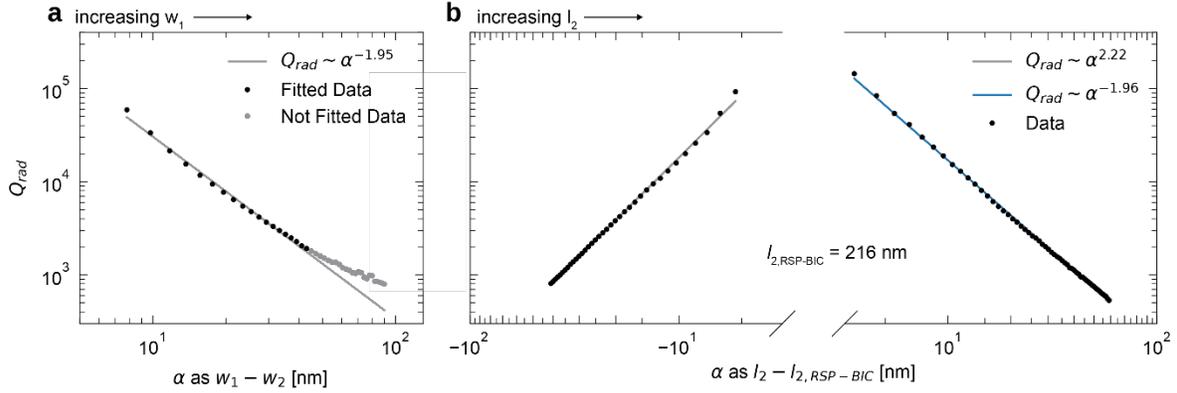

**Figure S1: Relation of Radiative Loss and Asymmetry.** This figure revisits the simulated $Q_{rad}$ from Figure 2e, with transmittance spectra for (a) varying $w_1$ and (b) varying $l_2$. Instead of using $w_1$ and $l_2$ as the x-axis, we define and use two asymmetry factors α as the x-axis. In (a), representing the $w_1$ sweep, we defined α as the difference between $w_1$ and $w_2$, where α = 0 corresponds to $w_1 = w_2 = 95$ nm, and α increases as $w_1$ increases. To reveal the SP-BIC's typical $Q_{rad} \sim \alpha^{-2}$ dependence, both axes are plotted on a logarithmic scale, which results in a linear relationship with the slope representing the exponent of the power law. As expected for this well-known geometry, the fitted $Q_{rad}$ values (black dots) lie on a straight line for smaller asymmetries. The gray curve is a fit excluding high asymmetries (gray dots), where the exponential trend breaks down, suggesting that α is no longer a suitable asymmetry factor in this range. The fit yields an exponent of -1.95, close to the theoretical value of -2. (b) Corresponding plots for the $l_2$ sweep from Figure 2e. Here, we define the symmetric length as the RSP-BIC length, $l_{2,\text{RSP-BIC}} = 216$ nm, and define α as $l_2$-$l_{2,\text{RSP-BIC}}$. In (b) the data is separated into two sets: α < 0 and α > 0. The region around α = 0 is cropped for clarity. In both regions, $Q_{rad}$ shows a nearly perfect linear dependence on $|\alpha|^2$, confirming the power law relationship. On the left, the fitted exponent of -2.22 is close to the theoretical value, while the exponent of -1.96 on the right side closely matches the theoretical value of -2, demonstrating that $Q_{rad}$ behaves similarly around the SP-BIC and RSP-BIC conditions.

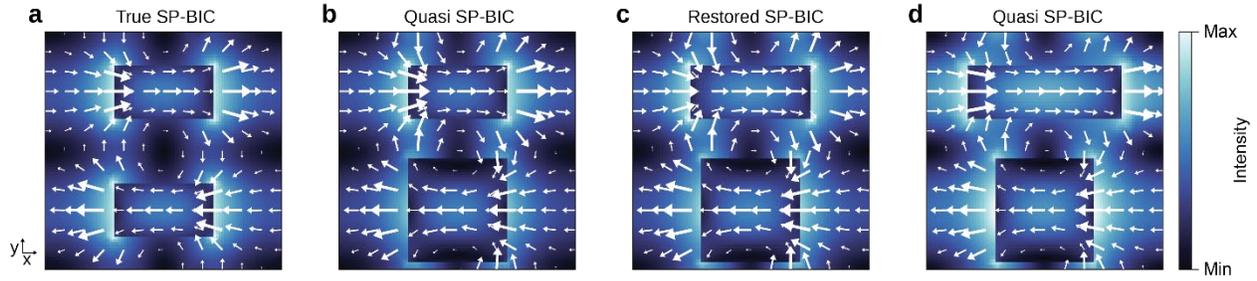

**Figure S2: Electric Field Profile of the SP-BIC Mode.** Electric field intensity profiles ($|E|^2$) within the x-y plane at $z = 57.5$ nm (center height) of the unit cell for different resonance conditions at their specific resonance wavelengths (see Figure 2d). The field intensity is color-coded from minimum to maximum within each individual cut. The direction of the electric field is represented by white arrows, with the arrow size indicating the field strength. (a) The true SP-BIC condition, structurally and optically symmetric, featuring two antiparallel dipole modes, one in each rod. (b) The quasi-SP-BIC condition, reached by increasing $w_1$, breaking the structural symmetry. (c) The RSP-BIC condition, obtained by further increasing $l_2$, restoring the optical symmetry. (d) The quasi-BIC condition after further increasing $l_2$. All four cases display a similar antiparallel dipolar mode profile.

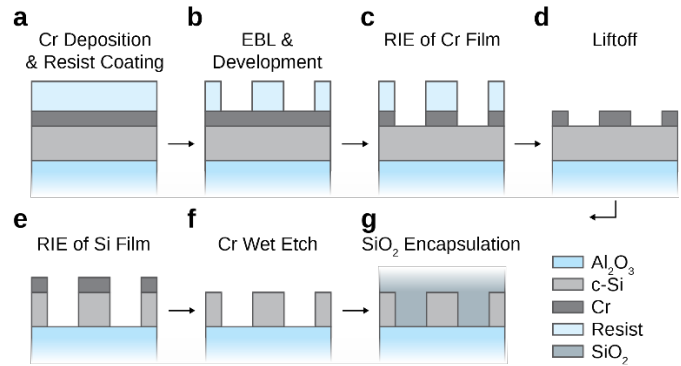

**Figure S3: Fabrication Workflow.** (a) The 115 nm crystalline silicon (c-Si) film on $Al_2O_3$ is first coated with a 40 nm layer of Cr via sputtering, followed by spin-coating with 125 nm of electron beam lithography (EBL) resist. (b) The resist is exposed in an EBL machine and developed using a two-step process. (c) The patterned resist serves as a mask for anisotropic etching of the Cr film using reactive ion etching (RIE). (d) During the liftoff the remaining resist is chemically removed. (e) The Cr layer is then used as a hard mask to etch down the 115 nm c-Si film. (f) The remaining Cr is chemically wet-etched. (g) Finally, the c-Si film is encapsulated in spin-on-glass (resembling $SiO_2$) with approximately 1000 nm height.

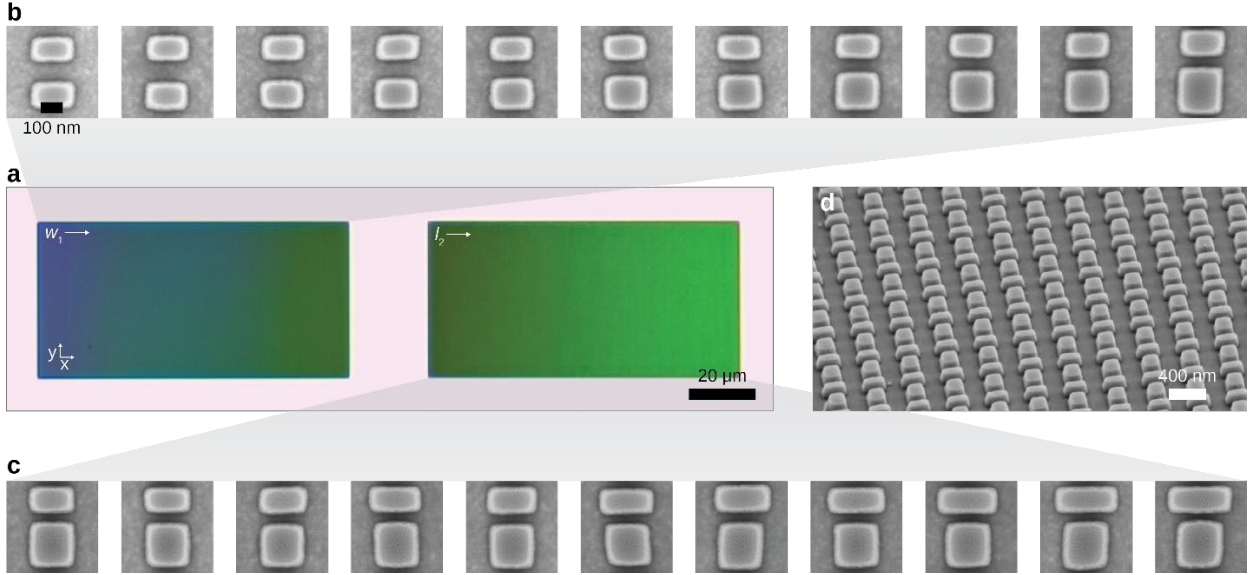

**Figure S4: Gradient Metasurfaces for Continuous Tuning of $\gamma_{rad}$.** (a) Optical image of the $w_1$ gradient (left) and the $l_2$ gradient (right). The $w_1$ gradient metasurface varies $w_1$ = 95-185 nm with the other parameters fixed as $p_x = p_y$ = 420 nm, $l_1 = l_2$ = 175 nm, and $w_2$ = 95 nm. The $l_2$ gradient metasurface varies $l_2$ = 175-280 nm with fixed $p_x = p_y$ = 420 nm, $l_1$ = 175 nm, and $w_1$ = 185 nm. Both metasurfaces are periodic along the y-direction, while $w_1$ and $l_2$ change smoothly along the x-direction, with an approximate step size of 0.4 nm between neighboring unit cells, as indicated by the gradual color change. (b) SEM images of individual unit cells taken at 10 μm intervals along the x-axis for the $w_1$ gradient, starting from 0 to 100 μm. (c) Equivalent SEM images along the $l_2$ gradient. (d) Angled SEM image of the $l_2$ gradient around the RSP-BIC condition.

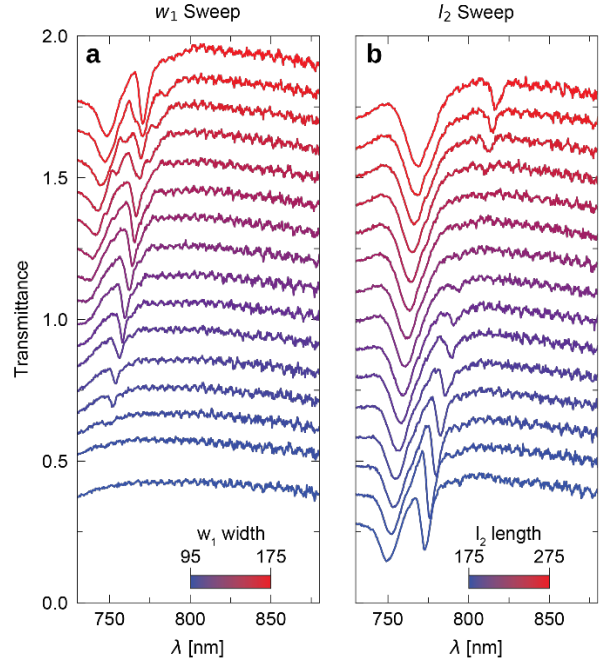

**Figure S5: Transmittance Spectra for $w_1$ and $l_2$ Gradient Metasurfaces.** (a) Transmittance spectrum for varying $w_1$ (equivalent to Figure 2h, left) indicated by the blue-to-red color map. The transmittance is offset by 0.1 for clarity. (b) Transmittance spectrum for varying $l_2$ (equivalent to Figure 2h, right), plotted similarly to (a). The white light used for taking the spectra was polarized along the x-axis of the sample.

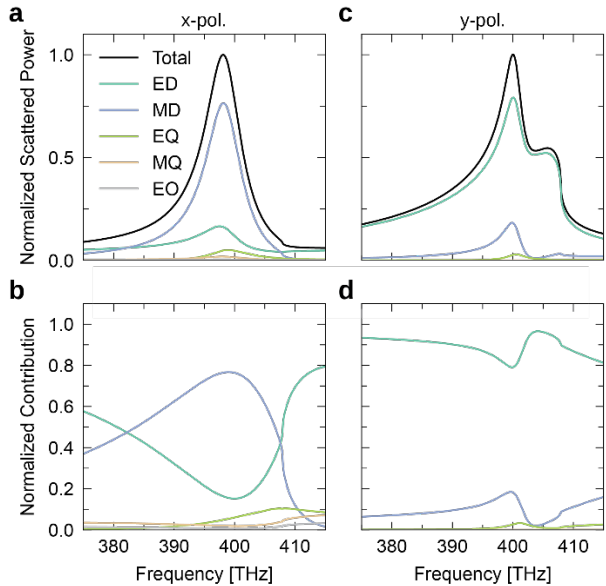

**Figure S6: Mode Decomposition.** Normalized scattered power of individual multipolar components for Rod 1 with $l_2 = 216$ nm for excitations in x-polarization (a) and y-polarization (c). (b), (d) Relative contributions $P_{multipole}/P_{total}$ to the total scattered power, demonstrating that the single resonance in x-polarization is dominated by the magnetic dipole MD, whereas the two resonances in y-polarization are mostly governed by the ED component. However, the left resonance at around 400 THz exhibits a dip in ED power and a peak in MD power, to which we assign the term MD-like ED-like to the right resonance at around 405 THz.

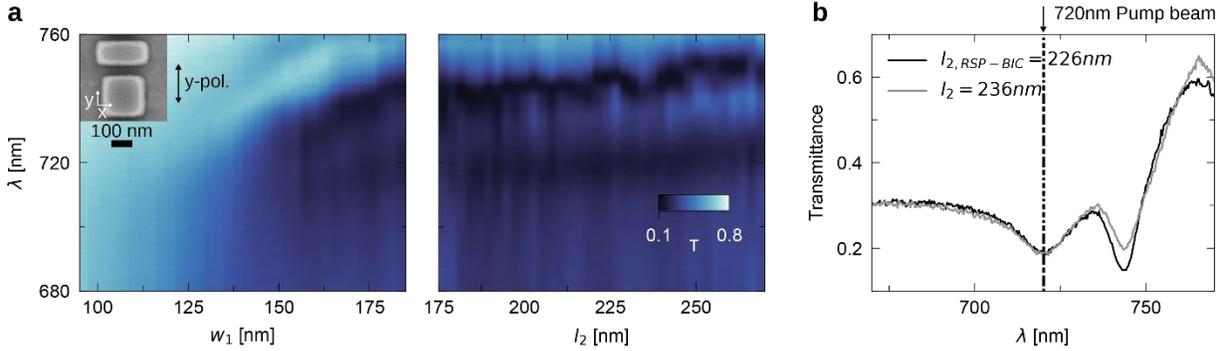

**Figure S7: Y-Polarized Transmittance Spectra for $w_1$ and $l_2$ Sweeps.** (a) Transmittance spectra for y-polarized light (inset for reference) in analogy to the x-polarized spectra of the two gradient metasurfaces in Figure 2h: on the left for varying $w_1$ with fixed $l_1 = l_2 = 175$ nm and $w_2 = 95$ nm; on the right for varying $l_2$ with fixed $w_1 = 185$ nm, $w_2 = 95$ nm, and $l_1 = 175$ nm. In the $w_1$ sweep (left), the Mie 1 and Mie 2 modes discussed in Figure 3 are not visible for small $w_1$ and appear as $w_1$ increases. In the $l_2$ sweep (right), there is almost no change in the transmittance spectra, indicating the mode's independence from the size parameters along the off-polarization axis. This allows us to use the same optimized pump wavelength of 720 nm, as it remains resonant with Mie 1 across the entire $l_2$ gradient. This is confirmed in (b) by the transmittance spectra at the RSP-BIC condition with $l_{2,\text{RSP-BIC}} = 226$ nm (shown in Figure 3a) and with $l_2 = 236$ nm (the structure used in Figures 3e,f), which both feature almost identical Mie 1 modes.

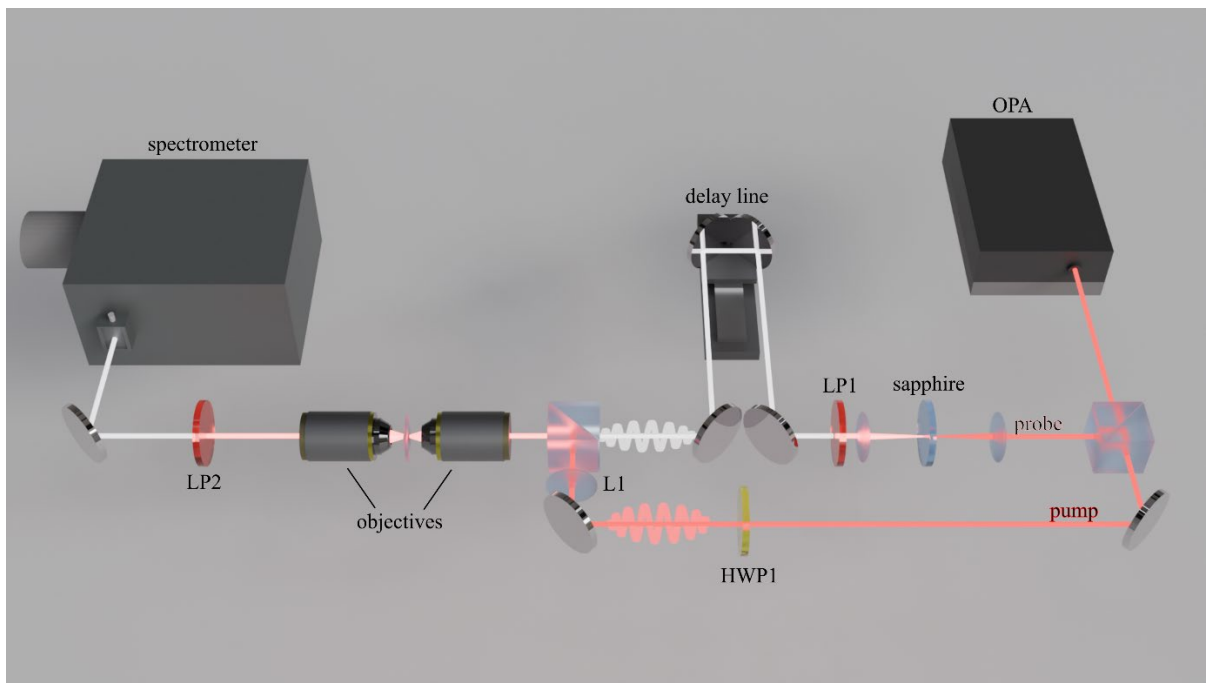

**Figure S8: Pump-Probe Setup.** The OPA output is split into a pump- and a probe path. The latter passes a sapphire window for spectral broadening, as well as a delay line to control the probe pulses' arrival time with respect to the pump pulses. A 750 nm long pass (LP1) filters out the residual pump after the sapphire. The pump's polarization is controlled with a half-wave plate (HWP1), and passes an additional lens (L1) before being recombined with the probe, which focuses it on the back focal plane of the condenser objective (10x, 0.25 NA). Pump and probe are collected using a second objective with the same specifications. The transmitted signal is filtered with a second 750 nm longpass filter (LP2), and analyzed using a spectrometer with a silicon-based CCD camera.

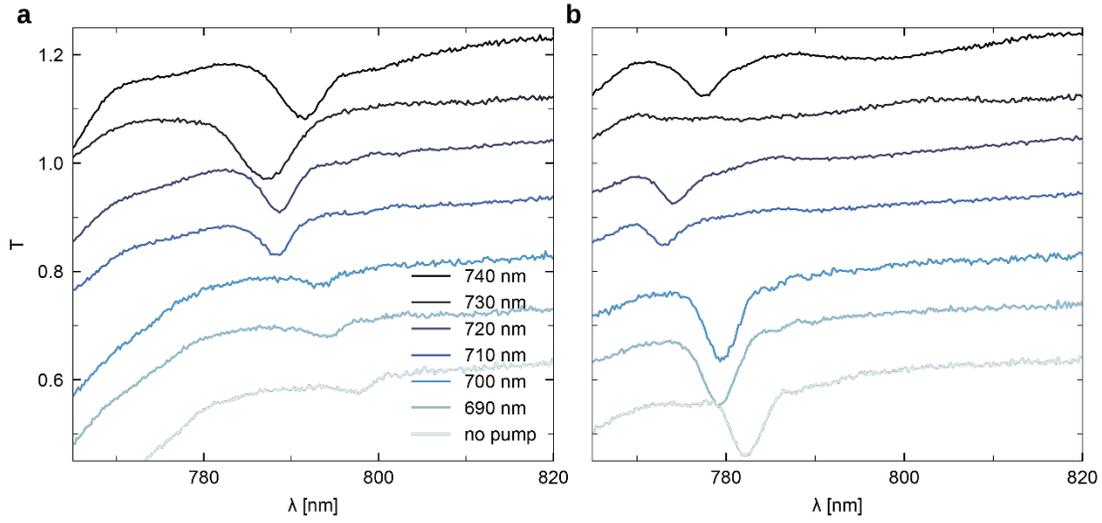

**Figure S9: Wavelength-dependent pumping.** Transmission spectra at different pump wavelengths at a fixed time delay of $t = 1$ ps and a fluence of 100 μJ/cm². (a) Resonance broadening position ($l_2 > l_{2,\text{RSP-BIC}}$) and (b) resonance sharpening position ($l_2 < l_{2,\text{RSP-BIC}}$). Consecutive spectra have been offset by 0.1 for clarity. Negligible changes in the resonance amplitude can be observed for pump wavelengths smaller than 710 nm, while the modulation reaches its maximum at 730 nm.

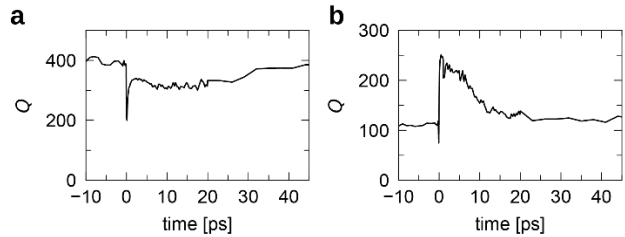

**Figure S10: Total Q-factors During Mode Broadening and Sharpening.** (a) Total Q-factor obtained using the TCMT model with variable $\gamma_{\text{rad}}$ and $\gamma_{\text{int}}$ for the transmittance time trace shown in Figure 4b with $l_2 = 236$ nm. The Q-factor decreases from an initial value of 400 down to 200 upon temporal overlap of the pump and probe pulses. It then stabilizes around 300 before gradually increasing back to its initial value of 400. (b) Equivalent representation to (a), but for $l_2 = 186$ nm and the time trace of Figure 4e. Upon the pump pulse, the Q-factor rises from an initial 110 to 250, then gradually converges back to the initial value within 20 ps.

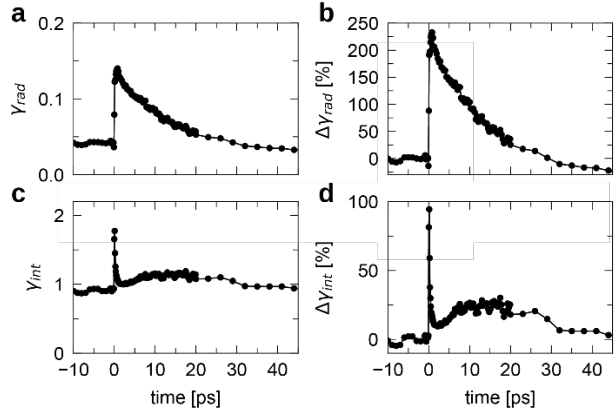

**Figure S11: Radiative and Intrinsic Loss Comparison.** For $l_2 = 236$ nm, the transmittance time trace is fitted using the TCMT model with variable $\gamma_{rad}$ and $\gamma_{int}$. (a) displays $\gamma_{rad}$, while (b) shows its relative change, defined as a percentage increase from the average $\gamma_{rad}$ before the pump. In (c), $\gamma_{int}$ and its corresponding relative values are shown. Notably, around $t = 0$ ps, coinciding with the arrival of the pump pulse, $\gamma_{int}$ experiences a sharp increase, suggesting a nonlinear interaction between the pump and probe beams. This spike subsides as the temporal overlap of the beams decreases. A sustained increase of approximately 25% persists over the full range of 0 to 30 ps relative to pre-pump values, substantially smaller than the peak increase of up to 250% observed for $\gamma_{rad}$.

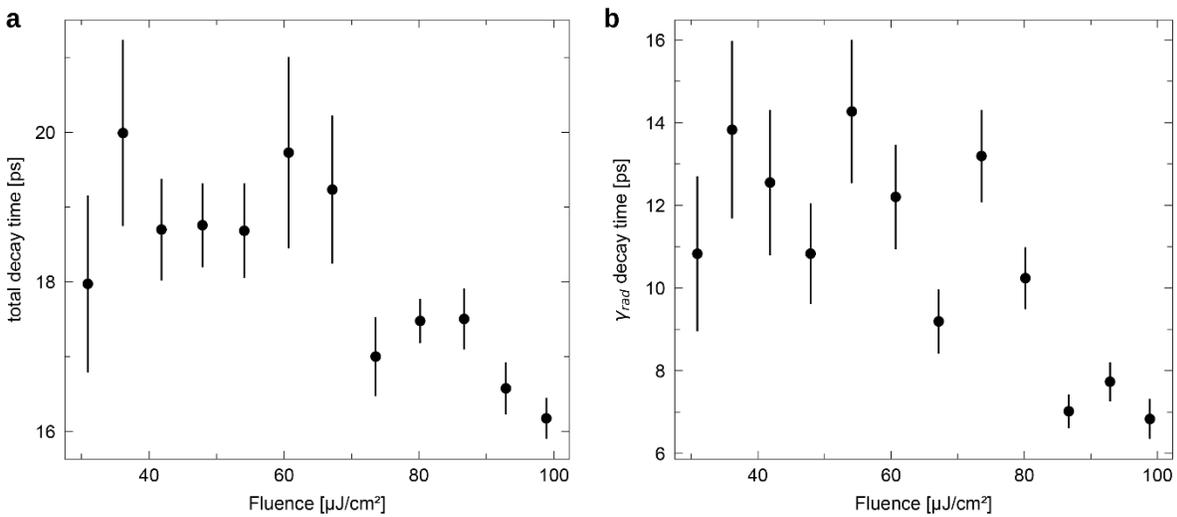

**Figure S12: Resonance Decay Times.** (a) 1/e decay time of the pump-induced spectral shift in dependence of the incident pump fluence. Decreasing decay times at high fluences suggest an increasing influence of Auger recombination. (b) 1/e decay time of the fitted radiative loss of the resonance. This time is generally shorter than the decay time of the spectral shift because the radiative loss depends quadratically on the asymmetry. Furthermore, because of Auger

recombination at high carrier densities, the excited populations in the two rods decay at different rates, causing their asymmetry to decay faster than the overall refractive index modulation.

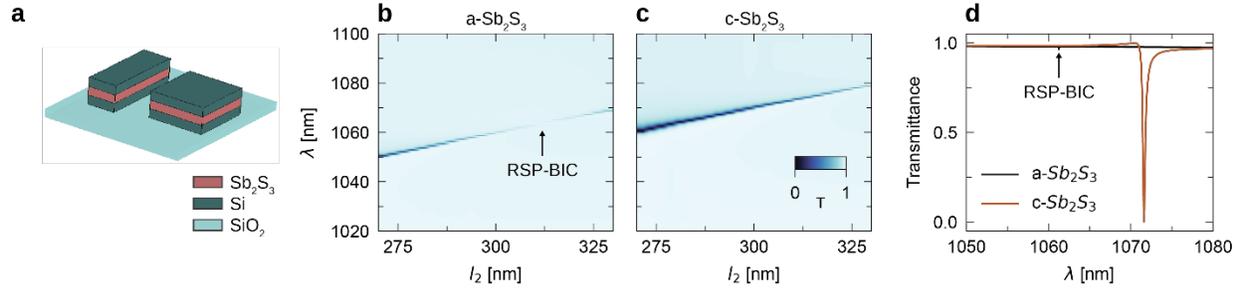

**Figure S13: Optically Induced Asymmetry Tuning in Sb$_2$S$_3$ Metasurfaces.** (a) Sketch of the unit cell, showing the 50 nm Sb$_2$S$_3$ layer at the center of the film stack. (b) Numerical transmittance spectra for amorphous Sb$_2$S$_3$ (a-Sb$_2$S$_3$) in both rods, with an $l_2$ sweep from 270 to 330 nm. The RSP-BIC condition is clearly observed around 1061 nm for $l_2 = 312$ nm. (c) Transmittance spectra for varying $l_2$, equivalent to (b), but with crystalline Sb$_2$S$_3$ (c-Sb$_2$S$_3$) in Rod 1. The RSP-BIC condition shifts significantly to higher $l_2$, indicating a strengthening of $p_1$ in Rod 1 compared to $p_2$ in Rod 2, and thus a change in $\gamma_{\text{rad}}$. (d) Transmittance spectra for the RSP-BIC condition are shown in (b) and (c) with $l_2 = 312$ nm in black and brown, respectively.